\documentclass[12pt]{article}

\usepackage{scicite}
\usepackage{graphicx}

\usepackage{newtxtext,newtxmath}
\newenvironment{sciabstract}{%
\begin{quote} \bf}
{\end{quote}}

\usepackage{xcolor}
\title{Inertial range of magnetorotational turbulence}

\author
{Yohei Kawazura$^{1,2,3\ast}$, Shigeo S. Kimura$^{1,4}$\\
\\
\normalsize{$^{1}$Frontier Research Institute for Interdisciplinary Sciences, Tohoku University}\\
\normalsize{6-3 Aoba, Aramaki, Sendai 980-8578, Japan}\\
\normalsize{$^{2}$Department of Geophysics, Graduate School of Science, Tohoku University}\\
\normalsize{6-3 Aoba, Aramaki, Sendai 980-8578, Japan}\\
\normalsize{$^{3}$School of Data Science and Management, Utsunomiya University}\\
\normalsize{350 Minemachi, Utsunomiya, Tochigi 321-8505 Japan}\\
\normalsize{$^{4}$Astronomical Institute, Tohoku University}\\
\normalsize{6-3 Aoba, Aramaki, Sendai 980-8578, Japan}\\
\\
\normalsize{$^\ast$To whom correspondence should be addressed; E-mail: kawazura@a.utsunomiya-u.ac.jp.}
}

\date{}

\begin{document} 


% \baselineskip24pt
\baselineskip18pt


\maketitle


\begin{sciabstract}
  Accretion disks around compact stars are formed due to turbulence driven by magnetorotational instability. 
  Despite over thirty years of numerous computational studies on magnetorotational turbulence, the properties of fluctuations in the inertial range --- where cross-scale energy transfer dominates over energy injection --- have remained elusive, primarily due to insufficient numerical resolution. 
  Here, we report the highest-resolution simulation of magnetorotational turbulence ever conducted.
  Our simulations reveal a constant cross-scale energy flux, a hallmark of the inertial range.
  We found that as the cascade proceeds to smaller scales in the inertial range, the kinetic and magnetic energies tend toward equipartitioning with the same spectral slope, and slow-magnetosonic fluctuations dominate over Alfv\'enic fluctuations, possessing twice the energy.
  These findings align remarkably with the theoretical expectations from the reduced magnetohydrodynamic model, which assumes a near-azimuthal mean magnetic field.
  Our results provide important implications for interpreting the radio observations by the Event Horizon Telescope.
\end{sciabstract}

% In setting up this template for *Science* papers, we've used both
% the \section* command and the \paragraph* command for topical
% divisions.  Which you use will of course depend on the type of paper
% you're writing.  Review Articles tend to have displayed headings, for
% which \section* is more appropriate; Research Articles, when they have
% formal topical divisions at all, tend to signal them with bold text
% that runs into the paragraph, for which \paragraph* is the right
% choice.  Either way, use the asterisk (*) modifier, as shown, to

\section*{Introduction}
Accretion disks around compact stars, such as black holes, neutron stars, and young stars, represent one of the most intriguing phenomena in astrophysics. 
For matter to accrete onto a compact star, the angular momentum of the matter must be transported outward.
It is widely believed that this angular momentum transport is achieved through turbulence driven by magnetorotational instability (MRI)~\cite{Balbus1991}.
Magnetorotational turbulence is such a rich process that it is not only crucial for angular momentum transport but also plays a critical role in the heating and acceleration of particles through dissipation of electromagnetic fluctuations~\cite{Quataert1998,Quataert1999,Kimura2016,Kimura2019,Sun2021}, and the energized particles are thought to be responsible for the observed emissions from these systems.
To understand particle energization, it is essential to elucidate the properties of turbulence in the inertial range~\cite{Stawarz2008,Schekochihin2009,Kawazura2019,Kawazura2020}, which bridges the energy injection scales and the dissipation scales.
In the inertial range, both energy injection and dissipation are subdominant compared to the cross-scale energy transfer by nonlinear effects.

While numerous studies numerically explored the turbulence driven by MRI over thirty years~\cite{Hawley1995,Fromang2007a,Fromang2007b,Lesur2011,Walker2016,Hirai2018}, the properties of fluctuations in the inertial range remain unknown.
For example, there is a clear discrepancy between the energy spectra in magnetorotational turbulence and those in the theoretical expectations of magnetohydrodynamic (MHD) turbulence~\cite{Goldreich1995,Boldyrev2006}. 
The discrepancy is attributed to the insufficient numerical resolution~\cite{Lesur2011}, although the statistical analysis of intermittent small-scale structures indicates that the spectra would eventually be consistent with those of MHD turbulence at a sufficiently high numerical resolution~\cite{Zhdankin2017b}.
In addition to the energy spectra, the energy partitioning of the MHD modes in the magnetorotational turbulence has not been investigated yet, whereas the partition has been numerically investigated in the MHD turbulence with artificial forcing~\cite{Cho2002,Cho2003,Makwana2020}.
The energy partition of the MHD modes is important for understanding the ion-to-electron heating ratio~\cite{Kawazura2020}, which is crucial for interpreting the radio observations by the Event Horizon Telescope.
The energy partition of the MHD modes also affects the particle acceleration efficiency in accretion flows~\cite{Teraki2019,Lynn2014}, which has a strong influence on the high-energy neutrino signals from nearby active galactic nuclei~\cite{Murase2020,Kimura2021,IceCube2022}.

Recently, these two mysteries in magnetorotational turbulence, namely, the energy spectra and the partition of MHD modes, were resolved using the reduced MHD model~\cite{Kawazura2022a}, which assumes the presence of a near-azimuthal mean magnetic field. 
More specifically, both the kinetic- and magnetic-energy spectra approach $k^{-3/2}$ with the same amplitude as the cascade proceeds where $k$ is the wavenumber, and the energy flux of the slow magnetosonic fluctuations is almost double that of the Alfv\'enic fluctuations.
This study aims to resolve the inertial range of magnetorotational turbulence by leveraging the power of the world's fastest supercomputer, and to examine whether the predictions made by the reduced MHD are valid or not.

\section*{Results}
Here, we present the direct numerical simulation of magnetorotational turbulence with the highest resolution in history.
The simulation was performed on the Fugaku supercomputer (the world's fastest machine until May 2022), using approximately 128 million central processing unit (CPU) hours.
We solved the incompressible MHD equations using a pseudospectral method in a local shearing box with $N_x\times N_y\times N_z = 8192\times 8192\times 4096$ grid points, where $x, y$, and $z$ denote radial, azimuthal, and vertical directions, respectively. 
The initial magnetic field was set to be vertically uniform with small-amplitude perturbations.
The size of the box was set to $L_x\times L_y\times L_z = 4\lambda\times 8\lambda\times 2\lambda$, where $\lambda = 2\pi v_\mathrm{A}/\Omega$ approximately equals the wavelength of the fastest-growing modes of MRI, $v_\mathrm{A}$ is the Alfv\'en speed given by the initial magnetic field, and $\Omega$ is the angular velocity of the accretion disk.
We note that the size of our simulation box is relatively smaller than those used in other simulations.
This configuration was chosen because our objective is to investigate the inertial range, which is expected to appear at scales smaller than $\lambda$.
A limitation of using a smaller box size is the inability to estimate the saturation amplitude of fluctuations and angular momentum transport, $\alpha$.
Therefore, we do not discuss these issues in this paper.
To narrow the dissipation range, the cascade was terminated by fourth-order hyperviscosity and hyperresistivity.
The viscous and resistive coefficients are set to the same value (i.e., magnetic Prandtl number Pm is unity).
Further details of the numerical setup can be found in Methods.

\begin{figure}
  \begin{center}
    \includegraphics*[width=1.0\textwidth]{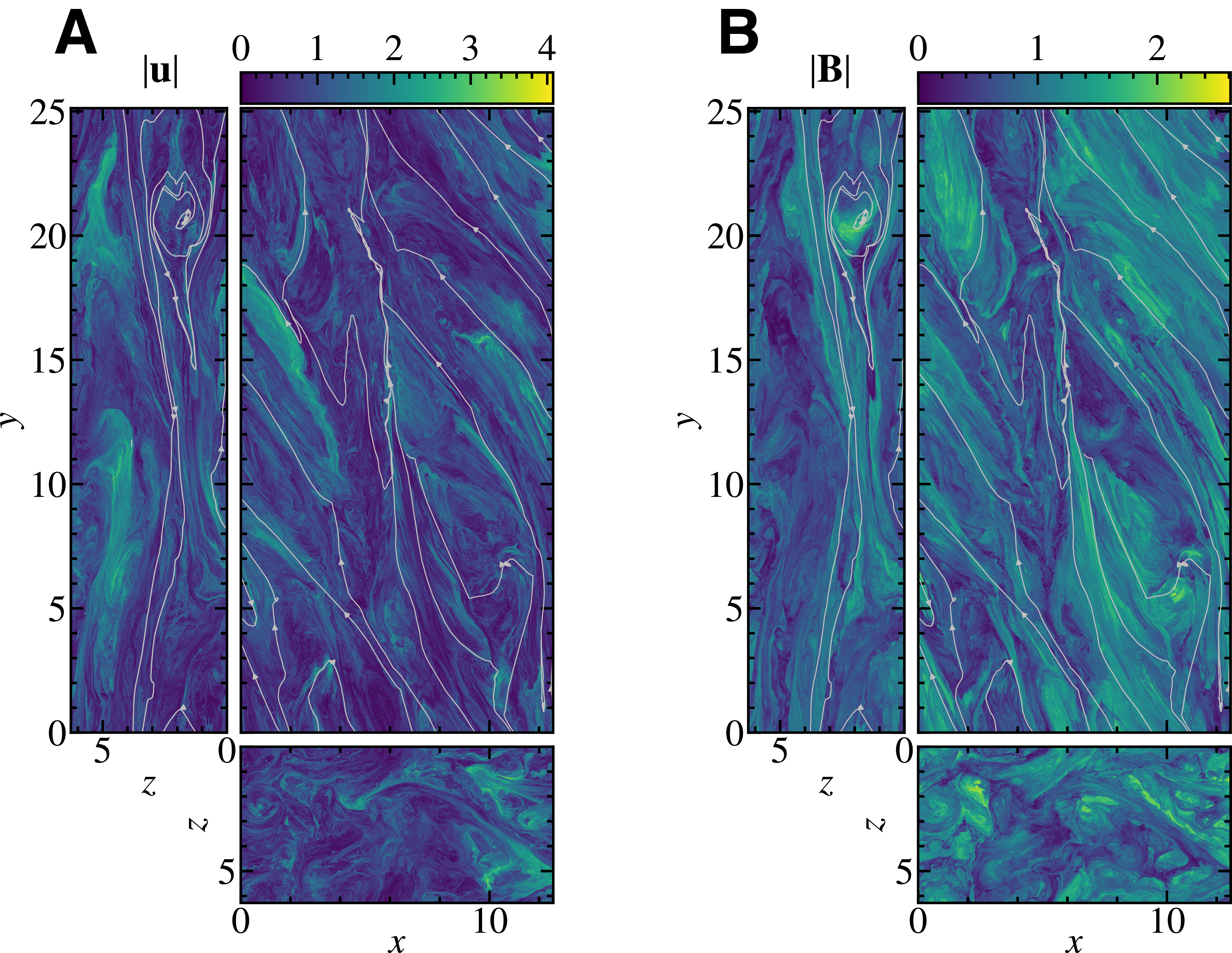}
  \end{center}
  \caption{
    \textbf{Morphology of finely resolved spatial structures of magnetorotational turbulence.} 
    Spatial distributions of \textbf{(A)} flow and \textbf{(B)} magnetic field intensity on $x = 0$, $y = 0$, and $z = 0$ planes, where $x, y$, and $z$ denote radial, azimuthal, and vertical directions, respectively. 
    The white lines are typical magnetic field lines. 
  }
  \label{f:snapshot}
\end{figure}

\begin{figure}
  \begin{center}
    \includegraphics*[width=1.0\textwidth]{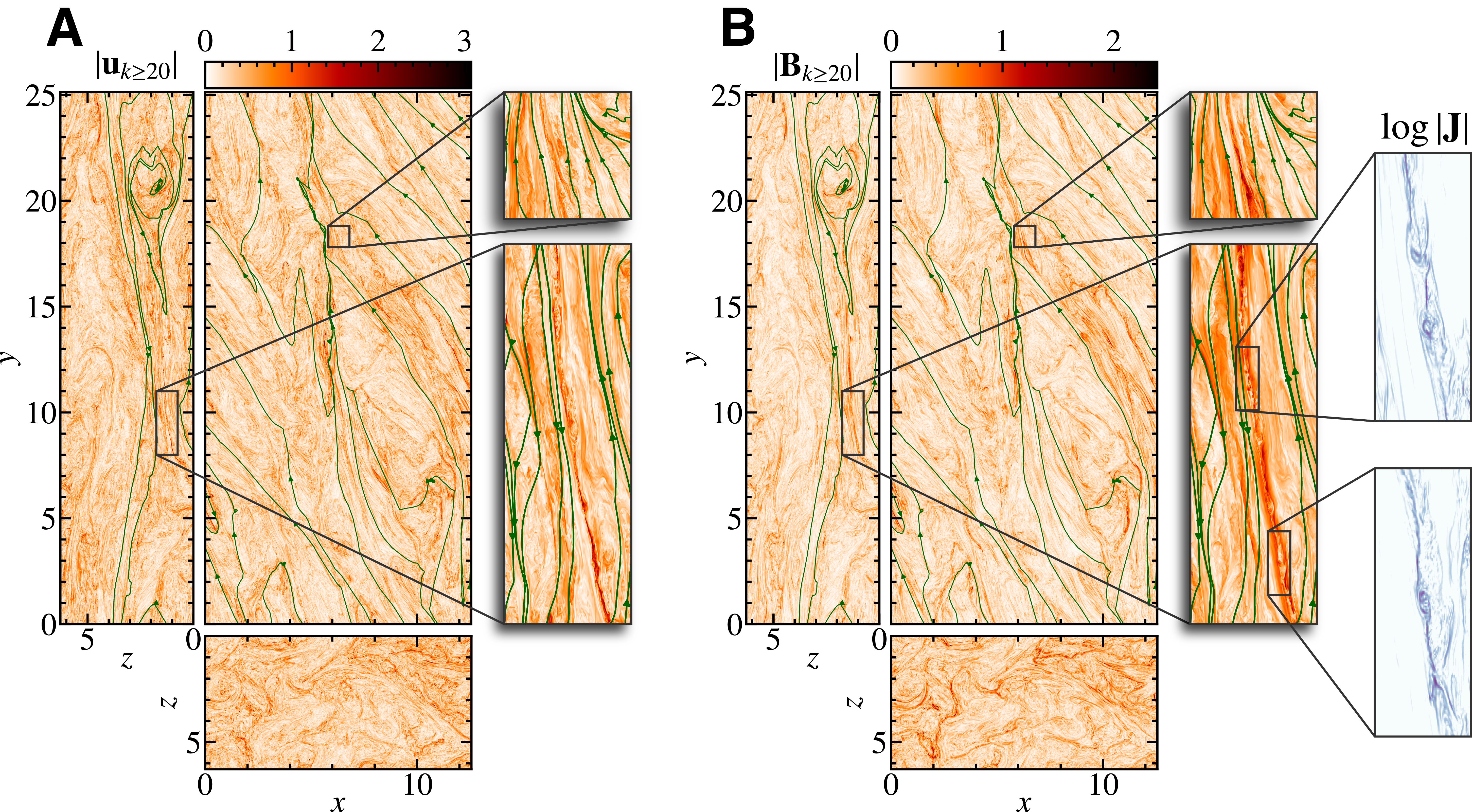}
  \end{center}
  \caption{
    \textbf{Small-scale structure of fluctuations.} 
    The same snapshot as Fig.~\ref{f:snapshot}, but high-pass filtered at $|\mathbf{k}| \ge 20$ (as shown below, the fluctuations with $|\mathbf{k}| \ge 20$ are in the inertial range).
    \textbf{(A)} and \textbf{(B)} correspond to the flow and magnetic field intensity distributions, respectively,
    The bluish pseudo color in \textbf{(B)} indicates the magnitude of unfiltered electric current density $\mathbf{J}$ in a logarithmic scale.
    The green lines are typical unfiltered magnetic field lines. 
  }
  \label{f:snapshot-filtered}
\end{figure}

\paragraph*{Morphology of fluctuations}
Figure~\ref{f:snapshot}, A and B show the snapshot of the norm of the flow field $\mathbf{u}$ and the magnetic field $\mathbf{B}$ together with typical magnetic field lines, in the planes $x = 0$, $y = 0$, and $z = 0$.
Each field is normalized by its root mean square value.
The overall structures of both fields are azimuthally elongated due to the Keplerian shear flow.
Upon closer inspection, the magnetic field has broad structures, while the flow field is concentrated in relatively smaller patches.
The reason for this difference is that the large-scale azimuthal magnetic field $B_y$ is preferentially amplified by the $\Omega$ effect~\cite{Walker2016}, which is evident from the shape of magnetic field lines on the $x = 0$ and $z = 0$ planes (see also Fig.~S1 for the snapshot of all components of $\mathbf{u}$ and $\mathbf{B}$ showing that only $B_y$ has intense large-scale structures).
One also notices that the magnetic field lines are predominantly azimuthal but have a finite radial component with anticorrelation between $B_x$ and $B_y$ (consistent with the hybrid PIC simulation of the shearing box magnetorotational turbulence~\cite{Kunz2016}; see also Fig.~S1, A and B).
Figure~\ref{f:snapshot-filtered}, A and B show the same snapshot of $\mathbf{u}$ and $\mathbf{B}$ but high-pass filtered by removing the fluctuation with a wavenumber smaller than 20.
As we will show later, the fluctuations with the wavenumber greater than $10$ are arguably in the inertial range.
The filtered $\mathbf{u}$ and $\mathbf{B}$ exhibit similar morphology, unlike their unfiltered counterparts, suggesting that the cascade tends to be Alfv\'enic as it moves towards smaller scales.
It is also observed that the spatial structures of the filtered fields are elongated along the magnetic field lines.
The magnified regions in Fig~\ref{f:snapshot-filtered}, A and B clearly visualize this elongation.
Thus, the large-scale magnetic field effectively acts like a mean field for the fluctuations in the inertial range.
One also finds that the azimuthal polarity of the mean magnetic field is not uniform, which is evident from the typical magnetic field lines plotted in Fig.~\ref{f:snapshot} and \ref{f:snapshot-filtered}.
This is because MRI creates both positive and negative $B_x$, which turn into positive and negative $B_y$ via shear flow.
Therefore, the emergence of the current sheet and magnetic reconnection of a mean field are naturally expected to occur in the simulation, and indeed, we find the chain of multiple plasmoids in the magnified region on the $x = 0$ plane in Fig.~\ref{f:snapshot-filtered}B. 
Nevertheless, these plasmoids do not fill a substantial portion of the simulation domain but appear only in limited regions.

\begin{figure}
  \begin{center}
    \includegraphics*[width=1.0\textwidth]{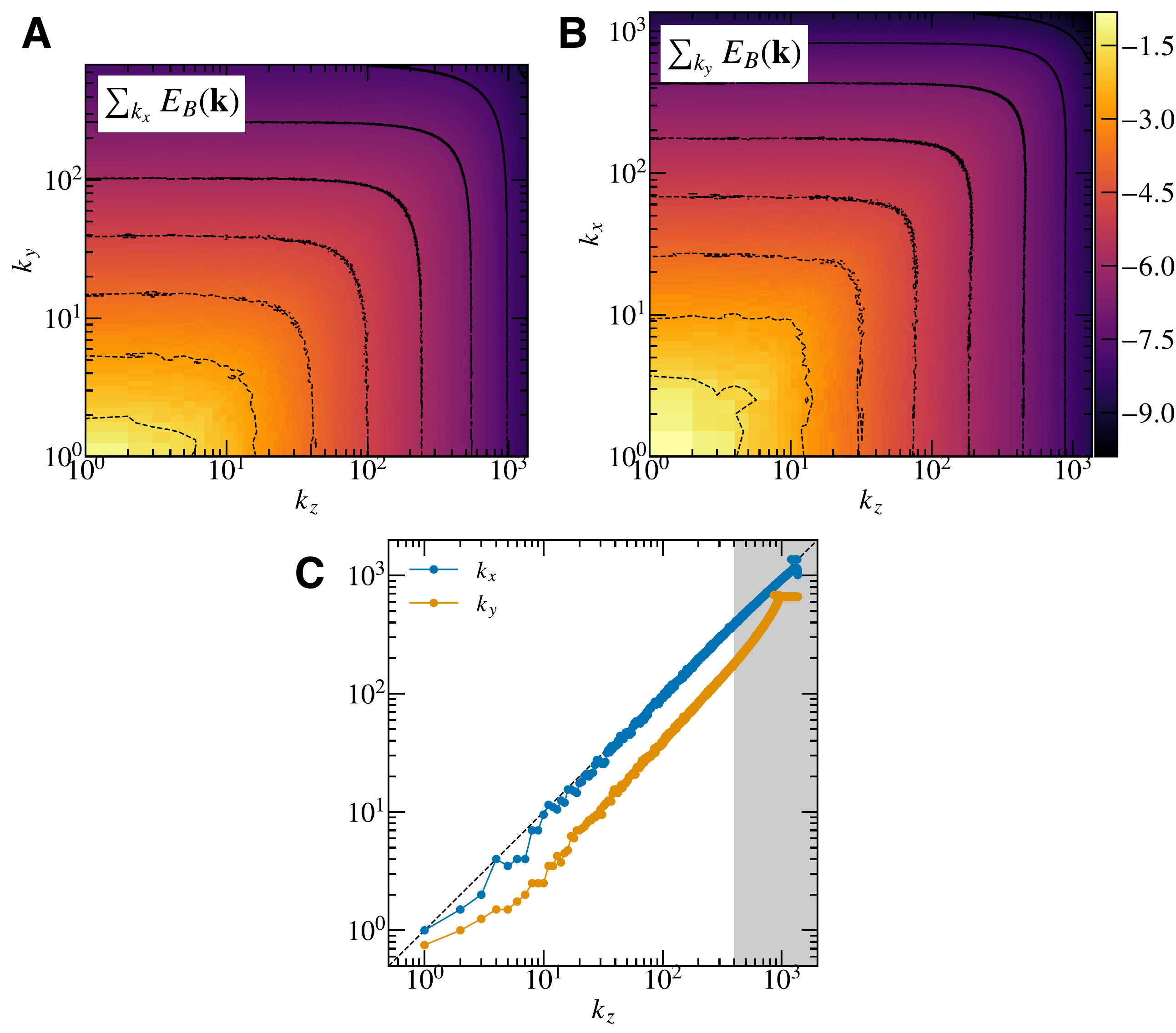}
  \end{center}
  \caption{
    \textbf{Wavenumber anisotropy of magnetic-energy spectrum.} 
    Two dimensional contour of magnetic-energy spectrum on \textbf{(A)} $(k_z, k_y)$ plane integrated over $k_x$ and on \textbf{(B)} $(k_z, k_x)$ plane integrated over $k_y$.
    The snapshot was taken immediately after remapping from the shearing coordinate to the laboratory coordinate so that the radial wavenumber $k_x$ in the both coordinate coincides.
    The colorbar applies to both \textbf{(A)} and \textbf{(B)}.
    \textbf{(C)} $k_z$-intercept vs. $k_y$- and $k_x$- intercepts of contour lines in \textbf{(A)} and \textbf{(B)}.
    The gray shaded area indicates to the dissipation range.
  }
  \label{f:2D spectra}
\end{figure}

\paragraph*{Spectral properties}
To explore the properties of small-scale fluctuations, we analyzed the energy spectra of $\mathbf{u}$ and $\mathbf{B}$. 
Figure~\ref{f:2D spectra}A and B present the two-dimensional spectrum of magnetic energy. 
Figure~\ref{f:2D spectra}A shows the spectrum as a function of $k_z$ and $k_y$, with the $k_x$-direction integrated out, while Fig.~\ref{f:2D spectra}B displays the spectrum as a function of $k_z$ and $k_x$, integrating out the $k_y$-direction.
The snapshot was taken immediately after remapping from the shearing coordinate to the laboratory coordinate, so that the radial wavenumber $k_x$ in both coordinates coincides (see Methods for details about the periodic remapping and the time dependency of $k_x$).
The results indicate anisotropy, specifically $k_x \simeq k_z > k_y$. 
This aligns with the observation seen in Figs.~\ref{f:snapshot} and \ref{f:snapshot-filtered} that the fluctuations are elongated along the shear and the mean magnetic field on the $x-y$ and $y-z$ planes, while the structures are nearly isotropic on the $x-z$ plane. 
However, Figure~\ref{f:2D spectra}C, which displays $k_x$ and $k_z$ as functions of $k_z$ by flattening Figures~\ref{f:2D spectra}A and B, shows that the anisotropy is scale-independent, unlike the scale-dependent anisotropy commonly seen in other simulations of MHD turbulence with external forcing and mean magnetic field. 
This is arguably because the mean magnetic field is not exactly $B_y$, and the perpendicular component of the wavenumber contaminates $k_y$.

\begin{figure}
  \begin{center}
    \includegraphics*[width=0.63\textwidth]{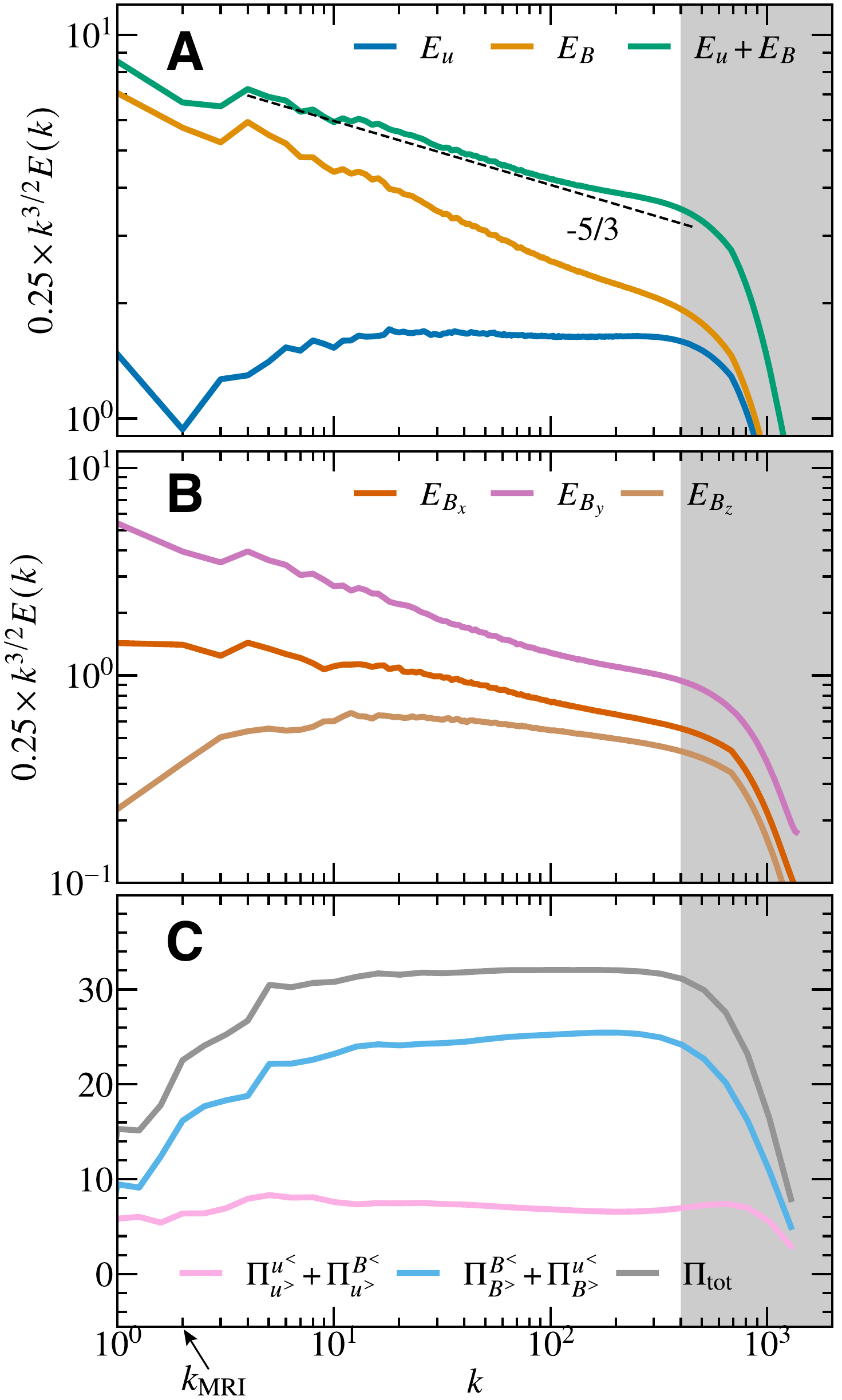}
  \end{center}
  \caption{
    \textbf{Spectra of magnetorotational turbulence.} 
    \textbf{(A)} Kinetic, magnetic, and total energy spectra compensated by $k^{3/2}$, where $k$ is the wavenumber.
    \textbf{(B)} Spectra of $x$, $y$, and $z$ components of magnetic energy.
    \textbf{(C)} Cross-scale energy flux $\Pi^{f^<}_{g^{>}}$ denoting the transfer from the field $f$ (which is the flow field when $f = u$ and is the magnetic field when $f = B$) with the wavenumber smaller than $k$ to the field $g$ with the wavenumber larger than $k$.
    The gray shaded area indicates to the dissipation range. The arrow indicates the wavenumber $k_\mathrm{MRI} = 2$ which denotes the wavenumber of the fastest growing MRI modes.
  }
  \label{f:spectra}
\end{figure}

Figure~\ref{f:spectra}A shows the omnidirectional spectra of kinetic, magnetic, and total energy.
The total energy spectrum is well fitted by a power law of $k^{-5/3}$ when $k\lesssim 100$, where $k$ is the wavenumber, and the spectral slope gets slightly shallower when $k\gtrsim 100$.
While the $-5/3$ spectrum was captured in the previously highest-resolution simulation~\cite{Walker2016}, the shallowing of the slope is found in this study for the first time.
We found that the shallowing occurs only in the magnetic field, while the kinetic-energy spectrum is almost completely $k^{-3/2}$ throughout the wavenumber domain.
In the previously highest-resolution simulation~\cite{Walker2016}, the kinetic-energy spectrum was slightly shallower than $k^{-3/2}$.
This is likely due to insufficient numerical resolution because the kinetic-energy spectrum becomes shallower than $k^{-3/2}$ in our low-resolution simulation to be shown below. 

The shallowing of the magnetic-energy spectrum is not surprising for the following reason.
The simulation of MHD turbulence with external forcing indicates that both magnetic- and kinetic-energy spectra converge to $k^{-3/2}$ with the same magnitude~\cite{Boldyrev2011}.
Meanwhile, as the cascade proceeds, the eddy turnover time decreases, causing fluctuations to lose memory of their MRI origin and eventually align with the above prediction~\cite{Walker2016,Kawazura2022a}.
At large scales, magnetic fluctuations are greater than those of kinetic energy and have a steeper spectrum, whereas the kinetic-energy spectrum is already $k^{-3/2}$, as shown in Fig.~\ref{f:spectra}A.
Thus, the magnetic-energy spectrum must flatten in order to merge into the kinetic-energy spectrum.

We further investigate the magnetic-energy spectrum by plotting $x, y$, and $z$ components in Fig.~\ref{f:spectra}B.
As is consistent with the previous simulation~\cite{Walker2016}, $B_y$ is dominant on the large scale (due to amplification via the $\Omega$ effect) and diminishes quickly in the smaller scales.
However, unlike the previous simulation, the $B_y$ spectrum becomes shallower at $k\gtrsim 100$, at which the total energy spectrum starts to become shallower than $-5/3$.
The second dominant component on the large scale is $B_x$, which is presumably due to MRI.
Although one might doubt that the shallowing of the spectrum is the numerical roll-up due to the hyper-dissipation, we do not think that this is the case because of the following two reasons. 
First, only the $B_y$ spectrum shows shallowing, while the other fields do not.
Second, the shallowing disappears in our low-numerical-resolution simulation, as we will show below.
We will further investigate this shallowing of $B_y$ later in this paper.
Although these two points do not rule out the possibility of numerical roll-up completely, it is currently computationally impossible to validate the presence of shallowing using Laplacian dissipation.

Evidence that our simulation resolved the inertial range is provided in Fig.~\ref{f:spectra}C, which plots the cross-scale energy flux through the wavenumber shell $|\mathbf{k}| = k$. 
Here, the energy flux $\Pi^{f^<}_{g^{>}}(k)$ denotes the energy transfer from the field $f$, where $f = u$ (or $=B$) for the flow (or magnetic) field, with the wavenumber smaller than $k$ to the field $g$ with the wavenumber larger than $k$ (see Methods for the mathematical definition).
We find that the total energy flux is fairly constant at $k\gtrsim 10$, which means that the cascade at $k\gtrsim 10$ is in the inertial range.
We stress that, in the previous simulations of MRI turbulence, the cross-scale energy flux was not constant~\cite{Lesur2011}, and our simulations is the first time that found the constant energy flux.
Note, however, that not only the number of grid points, but the size of the simulation box and the aspect ratio of their simulation are different from ours.
We also find that the amount of energy flux ``from $u$ and $B$ to $B$'' dominates that ``from $u$ and $B$ to $u$'', which is consistent with the observation by Ref.~\cite{Lesur2011}. 

\begin{figure}
  \begin{center}
    \includegraphics*[width=1.0\textwidth]{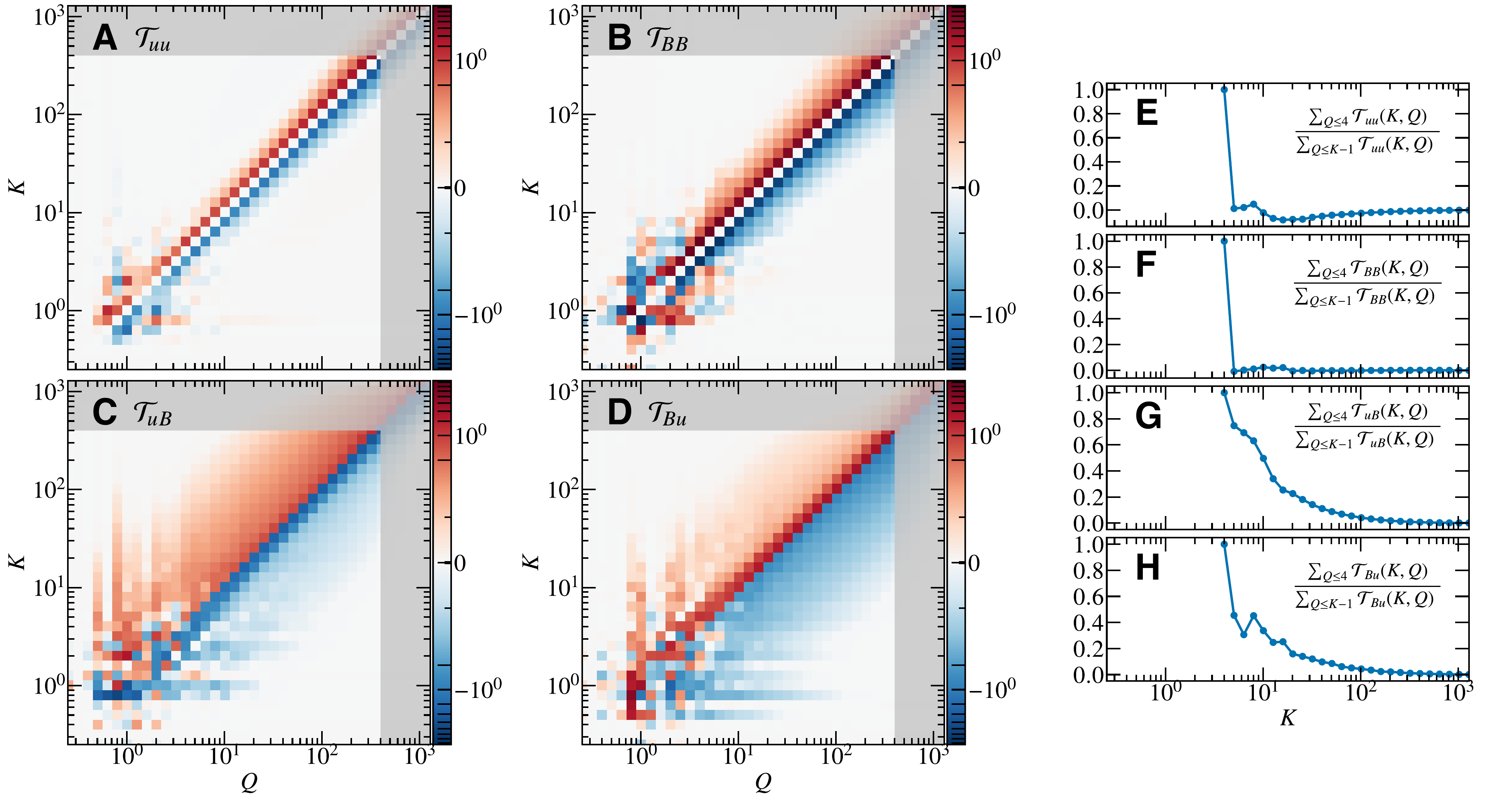}
  \end{center}
  \caption{
    \textbf{Locality of nonlinear energy transfer.}
    \textbf{(A)}-\textbf{(D)} Shell-to-Shell energy transfer function $\mathcal{T}_{fg}$, representing the transfer from field $f$ (the flow field for $f = u$ and the magnetic field for $f = B$) with wavenumber $Q$ to field $g$ with wavenumber $K$.
    \textbf{(E)}-\textbf{(H)} Contribution of the energy transfer from the injection range $k \le 4$.
  }
  \label{f:transfer}
\end{figure}

The nonlinear energy transfer is further investigated in Fig.~\ref{f:transfer}.
We define the transfer function $\mathcal{T}_{fg}(Q, K)$ that denotes the energy transfer from the field $f$ in the wavenumber shell $Q \le |\mathbf{k}| < Q+1$ to the field $g$ in the wavenumber shell $K \le |\mathbf{k}| < K + 1$~\cite{Verma2004,Debliquy2005,Alexakis2005,Grete2017} (see Methods for the mathematical definition).
We find that the dominant energy transfer among all possible combinations of fields is $\mathcal{T}_{BB}$, which is consistent with Ref.~\cite{Lesur2011}.
Noticeably, there is a transition at $|\mathbf{k}| \approx 4$.
In $|\mathbf{k}| > 4$, both $\mathcal{T}_{uu}$ and $\mathcal{T}_{BB}$ are fairly local, and the direction of the cascade is forward.
On the other hand, in $|\mathbf{k}| < 4$, none of the energy transfers are local.
We also find that there is an inverse energy transfer for $\mathcal{T}_{BB}$, suggesting that the large-scale structures of the magnetic field shown in Fig.~\ref{f:snapshot}A is formed not just by the $\Omega$ effect but also by the inverse cascade.
Furthermore, the transition scale, $|\mathbf{k}| \approx 4$, coincides with the scale where the energy flux becomes constant (Fig.~\ref{f:spectra}C), meaning that the energy cascade in the inertial range is local.
Regarding the energy transfers between $u$ and $B$, we find that they are more non-local than the transfer within $u$ or $B$ as $\mathcal{T}_{uB}$ and $\mathcal{T}_{Bu}$ have broad off-diagonal tails.
That being said, the non-local transfer from the injection range ($|\mathbf{k}| \le 4$) peters out as the cascade proceeds.
This is evident from Fig.~\ref{f:transfer}, E-H, which shows the contribution of energy transfer from the injection range defined by $\sum_{Q\le 4} \mathcal{T}_{fg}(K, Q)/\sum_{Q\le K - 1} \mathcal{T}_{fg}(K, Q)$.
We find that the contribution of the injection range in $\mathcal{T}_{uu}$ and $\mathcal{T}_{BB}$ disappears immediately below the transition scale $|\mathbf{k}| = 4$, which is obvious since these transfers are fairly local, as mentioned above. 
For $\mathcal{T}_{uB}$ and $\mathcal{T}_{Bu}$, the contribution of the injection range survives down to the relatively smaller scale, but the contribution becomes less than 10\% at $|\mathbf{k}| = 50$ for $\mathcal{T}_{uB}$ and at $|\mathbf{k}| = 40$ for $\mathcal{T}_{Bu}$.
This is dramatically different from Ref.~\cite{Lesur2011}, which showed that the box-scale to grid-scale transfer of $\mathcal{T}_{uB}$ was substantial. 
In short, the two facts that the cross-scale energy flux is constant and that the non-local energy transfer from the injection range peters out manifest that the cascade in our simulation is in the inertial range. This is one of the two main results of this paper.

\begin{figure}
  \begin{center}
    \includegraphics*[width=1.0\textwidth]{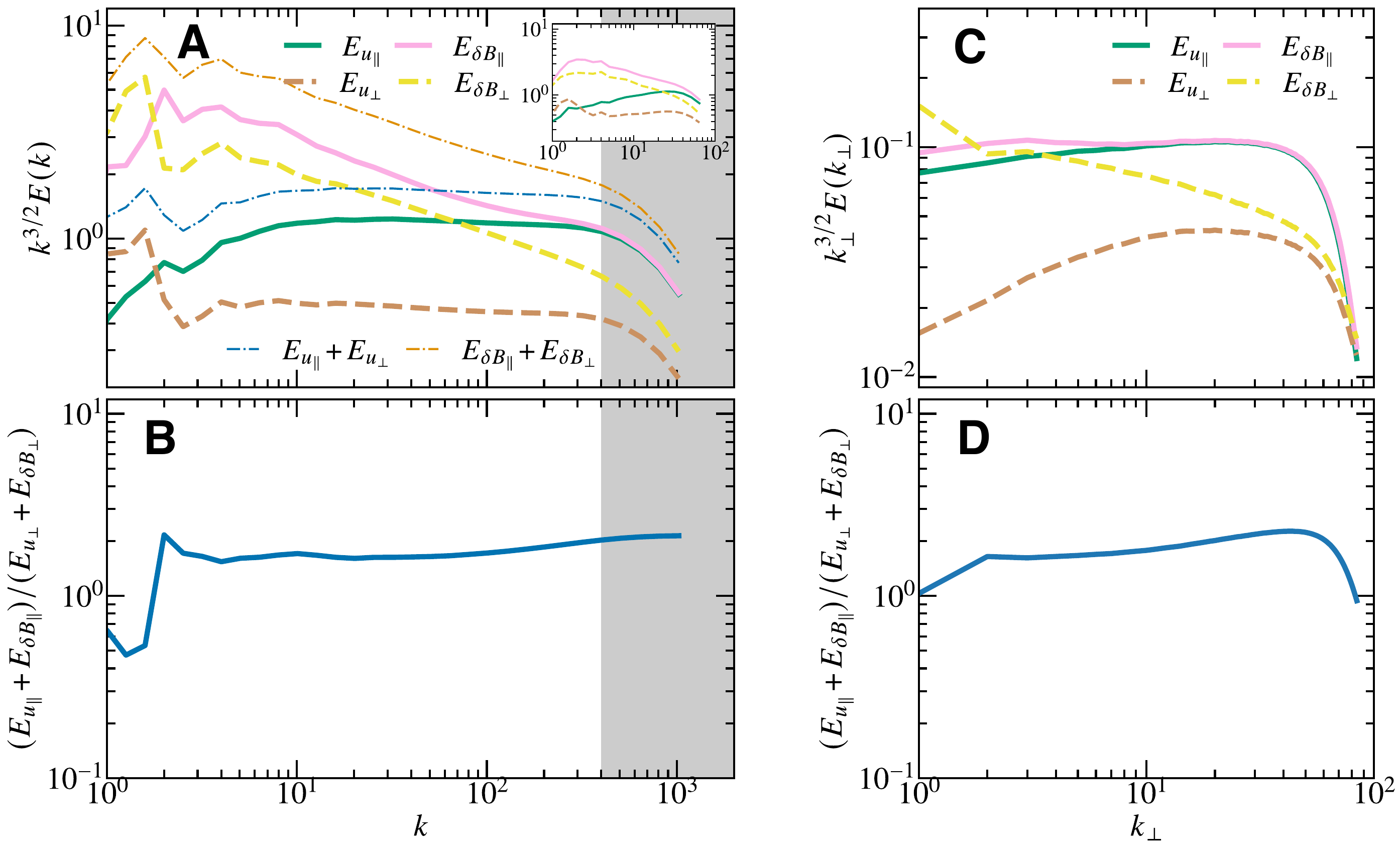}
  \end{center}
  \caption{
    \textbf{Decomposition of flow and magnetic fields into Alfv\'en and slow-magnetosonic fluctuations.}
    \textbf{(A)} and \textbf{(C)} spectra of Alfv\'en and slow-magnetosonic fluctuations, respectively.
    The inset in \textbf{(A)} shows the spectra obtained by the low resolution simulation with $N_x\times N_y\times N_z = 512\times 512\times 256$.
    \textbf{(B)} and \textbf{(D)} ratio of slow-magnetosonic to Alfv\'enic fluctuations.
    \textbf{(A)} and \textbf{(B)} are obtained from the MHD simulation presented in this paper, while \textbf{(C)} and \textbf{(D)} are obtained from the reduced MHD simulation~\cite{Kawazura2022a}.
  }
  \label{f:SW-AW}
\end{figure}

\paragraph*{Partition between slow-magnetosonic and Alfv\'enic fluctuations}
The other main result of this paper, namely, the partition between slow-magnetosonic and Alfv\'enic fluctuations in the magnetorotational turbulence, is shown in Fig.~\ref{f:SW-AW}. 
In incompressible MHD, the flow and magnetic field of the Alfv\'en waves are described by $u_\perp \hat{\mathbf{k}}_\| \times \hat{\mathbf{k}}_\perp$ and $\delta B_\perp \hat{\mathbf{k}}_\| \times \hat{\mathbf{k}}_\perp$ while those of the slow-magnetosonic waves are described by $u_\|[\hat{\mathbf{k}}_\| - (k_\|/k_\perp)\hat{\mathbf{k}}_\perp]$ and $\delta B_\|[\hat{\mathbf{k}}_\| - (k_\|/k_\perp)\hat{\mathbf{k}}_\perp]$, respectively, where the hat symbol denotes a unit vector, and the eigenfunction of the slow magnetosonic waves is used.
As we found in Figs.~\ref{f:snapshot} and \ref{f:2D spectra}, the wavenumber is strongly anisotropic, i.e., $k_\|/k_\perp \ll 1$.
This allows the slow-magnetosonic waves to be represented simply by $u_\|$ and $\delta B_\|$ (Fig.~S3 confirms that neglecting the terms proportional to $k_\|/k_\perp$ does not change the spectra).
Thus, we can decompose the total magnetic and kinetic energy into those of Alfv\'enic and slow-magnetosonic fluctuations via projection of $\mathbf{u}$ and $\mathbf{B}$ onto the mean magnetic field.
However, the global mean magnetic field does not always serve as a mean field for small-scale fluctuations, and thus, we use the method developed by Cho and Lazarian~\cite{Cho2004} to decompose the fluctuation and the local mean magnetic field $\mathbf{B}_0(\mathbf{r})$. 
For a given wavenumber $k$, the local mean magnetic field $\mathbf{B}_0(\mathbf{r})$ is obtained by filtering the Fourier modes of $\mathbf{B}$ with the wavenumber greater than $k/2$, and the fluctuations are obtained by filtering out the Fourier modes of $\mathbf{u}$ and $\mathbf{B}$ with the wavenumber smaller than $k/2$ or greater than $2k$.
Then we decompose the fluctuations of $\mathbf{u}$ and $\mathbf{B}$ into parallel ($u_\|$ and $\delta B_\|$) and perpendicular ($u_\perp$ and $\delta B_\perp$) components to $\mathbf{B}_0$.
Figure~\ref{f:SW-AW}A shows the spectra of the decomposed fields.
First, we confirm the validity of our decomposition by comparing the sum of $u_\|$ and $u_\perp$ with the total $u$ in Fig.~1A and the sum of $\delta B_\|$ and $\delta B_\perp$ with the total $B$ in Fig.~\ref{f:snapshot}A;
the spectra of the total fields are almost perfectly recovered from the sum of the decomposed fields.
In terms of magnetic fluctuations, one finds that only the spectrum of slow-magnetosonic fluctuations exhibits flattening at $k \gtrsim 100$, while that of Alfv\'enic fluctuations has nearly the same spectral index throughout the inertial range.
In contrast, the kinetic-energy spectra of both Alfv\'en and slow-magnetosonic fluctuations are almost perfectly $k^{-3/2}$.
Therefore, the flattening of the spectrum seen in Fig.~\ref{f:snapshot}A is solely due to the magnetic component of slow-magnetosonic fluctuations.
The inset of Fig.~\ref{f:SW-AW}A shows the same spectra obtained from the simulation with lower resolution $N_x\times N_y\times N_z = 512\times 512\times 256$, which manifests that the flattening of the magnetic energy of the slow-magnetosonic waves is absent.
Thus, the flattening can be seen only with ultra-high resolution.
It can be seen that due to this flattening, the magnetic energy and kinetic energy of slow-magnetosonic fluctuations tend to be equipartitioned, and both spectra approach $k^{-3/2}$.
On the other hand, the kinetic and magnetic-energy spectra of Alfv\'enic fluctuations do not converge in our simulation, and a further higher resolution is required to determine the converged spectral slope.
Figure~\ref{f:SW-AW}B shows the ratio of the energy of slow-magnetosonic fluctuations to that of Alfv\'enic fluctuations, manifesting that slow-magnetosonic fluctuations have approximately twice stronger energy than the Alfv\'enic ones.
One also finds that the ratio is almost constant throughout the wavenumber domain, suggesting that the coupling between the Alfv\'enic and slow-magnetosonic fluctuations in MRI turbulence is weak.
This is consistent with the recent report on the shearing box simulation which found that the ratio of energy injection between Alfv\'enic and slow-magnetosonic fluctuations almost equals to the ratio of dissipation which we computed using the reduced MHD~\cite{Satapathy2024}.

Finally, we compare our results with the simulation of magnetorotational turbulence solved by the reduced MHD with a near-azimuthal mean magnetic field~\cite{Kawazura2022a}.
Figures~\ref{f:SW-AW}C and D are reproduced from Ref.\cite{Kawazura2022a} and show qualitative consistency with our results in Figs.~\ref{f:SW-AW}A and B.
Specifically, Fig.~\ref{f:SW-AW}C displays the spectra of the slow-magnetosonic and Alfv\'enic fluctuations, which are remarkably similar to the spectra at $k \gtrsim 100$ in Figure~\ref{f:SW-AW}A.
Figure~\ref{f:SW-AW}D shows the ratio between the two fluctuations, with a value of $\approx 2$, exactly matching the result in Fig.~\ref{f:SW-AW}B.
Thus, we conclude that the reduced MHD with a near-azimuthal mean magnetic field effectively captures the features of the inertial range of magnetorotational turbulence.

\section*{Discussion}
The validity of reduced MHD in solving the inertial range of magnetorotational turbulence is now supported by the two findings of this study, namely (1) that the spatial structures of the small-scale fluctuations in our magnetorotational turbulence are elongated along the large-scale magnetic field which is azimuthally elongated and (2) that their spectral shapes remarkably resemble those obtained by reduced MHD.
Although the results of our MHD simulation are formally applicable when $\beta$, thermal-to-magnetic pressure ratio, is infinitely large as we solved incompressible MHD, the simulations of reduced MHD showed that the spectral shape and the energy partition between the slow-magnetosonic and Alfv\'enic fluctuations (viz., by the factor of two) do not depend on $\beta$.
Thus, we think that the results shown in this paper are also valid in the smaller $\beta$ regime.
Note that the recent study shows that the ratio of energy injection between the slow-magnetosonic and Alfv\'enic fluctuations does not depend on $\beta$~\cite{Satapathy2024}, which supports our claim.
Apart from $\beta$, our results can depend on the value of Pm and on the presence or absence of net magnetic flux. 
While we have explored only the case with Pm = 1, with finite vertical net flux, and without azimuthal net flux, numerous simulations have investigated other cases (e.g., see Ref.~\cite{Loren2024} for the latest study).
However, these previous parameter scans mostly focus on the dependence of the value of $\alpha$ and saturation amplitude.
Therefore, it would be very interesting to investigate whether the results of this study (e.g., the partition between slow-magnetosonic and Alfv\'enic fluctuations) change under different settings.

The energy partition between the slow-magnetosonic and Alfv\'enic fluctuations is important for inferring the ion vs. electron heating in hot accretion flows, such as at M87 and Sgr A*.
The ion-to-electron heating ratio is a key parameter for theoretical understanding of radio observations by the Event Horizon Telescope~\cite{EHT2019e}.
In collisionless magnetorotational turbulence, the half of the energy flux injected via MRI is supposed to be viscously dissipated due to the pressure anisotropy~\cite{Sharma2007,Kempski2019}, and the remaining half turns into ion and electron heating at the microscopic scales smaller than the ion Larmor radius. 
Our previous study of microscopic turbulence using hybrid gyrokinetics showed that, regarding the heating at the microscopic scales, ions are heated more efficiently than electrons when slow-mode-like compressive fluctuations dominate the Alfv\'enic ones~\cite{Kawazura2020}.
Thus, the fact that the slow-magnetosonic fluctuations have twice as large energy as the Alfv\'enic ones indicates that at least half of the energy cascaded down to the ion Larmor scale dissipates into ion heating (see also \cite{Kawazura2022a} for some caveats of this conclusion). 
Recently, the one dimensional energy transport model of hot accretion disks showed that the preferential ion heating due to the slow-mode-like compressive fluctuations can substantially influence the global temperature distribution~\cite{Satapathy2023}.
We should note that this discussion assumes that MRI turbulence is active in the accretion flows around M87 and Sgr A*~\cite{EHT2021,EHT2022}. 
However, the theoretical examinations of the Event Horizon Telescope data prefer magnetically arrested disk (MAD) regime, in which MRI is supposed to be strongly (but not completely) suppressed due to strong magnetic fields. 
That being said, the contribution of MRI-driven turbulence to the radiation profile in a MAD state is still being debated.
Furthermore, the accumulation and amplification of the poloidal magnetic field due to accretion caused by MRI are crucial for achieving the MAD state~\cite{Narayan2003, Jacquemin-Ide2023}.
There is also a study that suggests MRI is not even suppressed in MADs, contrary to previous claims~\cite{Begelman2022}.

MRI turbulence can accelerate high-energy non-thermal particles via magnetic reconnection~\cite{Hoshino2013,Hoshino2015,Kunz2016,Bacchini2022}, and these higher energy particles are further accelerated via stochastic acceleration through wave-particle interactions~\cite{Comisso2018}. 
If non-thermal protons are accelerated to higher energies, these protons can produce cosmic neutrinos seen by IceCube experiments~\cite{IceCube2022}.
Previous studies on stochastic acceleration using MHD and test-particle simulations exhibit particle acceleration with a hard-sphere type diffusion coefficient in momentum space \cite{Lynn2014,Kimura2019,Sun2021}. 
However, these MHD simulations lack sufficient spacial resolution, underestimating the acceleration efficiency at lower energies. Our highest resolution MRI simulation resolves the inertial range of MRI turbulence for the first time. 
This allows us to evaluate the particle acceleration efficiency at much lower energies down to dissipation scale. 
This would shed light on modeling particle acceleration inside accretion flows.

Lastly, as far as one aims to explore the inertial range of magnetorotational turbulence, extremely expensive full MHD simulations (like we did in this study) are not necessary, and simulations of reduced MHD with a small amount of computational cost would be enough.
More importantly, the adequacy of reduced MHD opens up the possibility of exploring magnetorotational turbulence in the collisionless regime where the MHD approximation formally breaks down.
There have been a number of numerical studies of collisionless magnetorotational turbulence (e.g.,~\cite{Sharma2006, Sharma2007,Riquelme2012,Hoshino2013,Hoshino2015,Kunz2016,Foucart2016,Foucart2017,Inchingolo2018,Kempski2019,Bacchini2022}), but presumably, the inertial range was not well resolved because the models used in these studies are much more complicated and numerically harder to solve than MHD.
However, this study suggests that it is possible to reach the inertial range of collisionless magnetorotational turbulence using the reduced kinetic MHD~\cite{Schekochihin2009} in a rotating frame, whose collisional limit is the reduced MHD which we used in \cite{Kawazura2022a}.

\section*{Methods}
\paragraph*{Governing equations}
We consider a local Cartesian coordinate that corotates with the accretion disk at a radial distance $r = r_0$ from the center.
The coordinate labels $(x, y, z)$ denote the radial, azimuthal, and vertical directions, respectively.
We solve the incompressible MHD equations in this coordinate system,
%----------------------------------------------------------------% 
\begin{align}
  \frac{\partial \mathbf{u}}{\partial t} + (\mathbf{u}_0 + \mathbf{u}) \cdot \nabla \mathbf{u} = -\nabla P + \mathbf{B} \cdot \nabla \mathbf{B} - 2 \Omega \hat{\mathbf{z}} \times \mathbf{u} - \mathbf{u} \cdot \nabla \mathbf{u}_{0} \
  \label{e:MHD_INCOMP u}\\
  \frac{\partial \mathbf{B}}{\partial t} + (\mathbf{u}_{0} + \mathbf{u}) \cdot \nabla \mathbf{B} = \mathbf{B} \cdot \nabla (\mathbf{u}_{0} + \mathbf{u}) \
  \label{e:MHD_INCOMP B}\\
  \nabla \cdot \mathbf{u} = 0,\; \nabla \cdot \mathbf{B}=0,
  \label{e:MHD_INCOMP div u div B}
\end{align}
%----------------------------------------------------------------% 
where $\mathbf{u}$ is the flow velocity, $\mathbf{B}$ is the magnetic field, $P$ is the total pressure, $\Omega$ is the local angular velocity of the disk, $q = -(\mathrm{d}\ln \Omega/\mathrm{d} \ln r)_{r=r_0}$ is the shear rate, and $\mathbf{u}_0 = -q\Omega x \hat{\mathbf{y}}$ is the background flow.
Due to the incompressible condition, the density $\rho$ is spatio-temporally constant.
Although MRI turbulence can excite substantial acoustic wave power in accretion flows, the MRI turbulence itself is highly incompressible.
In fact, the particle-in-cell simulation demonstrated that the incompressible approximation is valid in collisionless magnetorotational turbulence in a shearing box~\cite{Bacchini2024}. 
We assume that the rotation of the disk is Keplerian, i.e., $q = 3/2$.
The boundary condition is set to periodic in $y$ and $z$, and to shearing periodic in $x$~\cite{Hawley1995}.

\paragraph*{Numerical setup}
We numerically solve \eqref{e:MHD_INCOMP u}-\eqref{e:MHD_INCOMP div u div B} via pseudo-spectral code \textsc{\textsf{Calliope}}~\cite{Kawazura2022b,Kawazura2022d}.
In order to adopt the pseudospectral method, we enforce the triply periodic boundary conditions by transforming to the shearing coordinate, $y \mapsto y - q\Omega t x$.
This transformation makes the radial wavenumber time dependent, $k_x(t) = k_x + q\Omega t k_y$, where $k_x$ and $k_x(t)$ are the radial wavenumber in the shearing frame and the laboratory frame, respectively.
To avoid $k_x$ from ever growing, we adopt the remapping method where the fields are mapped to the original non-shearing coordinate every $T = L_y/(q\Omega L_x)$~\cite{Rogallo1981,Umurhan2004}.

We set the initial magnetic field as the sum of the uniform vertical field $B_0\hat{\mathbf{z}}$ and the random fluctuations with amplitude much smaller than $B_0$.
The size and aspect ratio of the computational domain are set to $L_x\times L_y\times L_z = 4\lambda\times 8\lambda\times 2\lambda$, where $\lambda = 2\pi v_\mathrm{A}/\Omega$ approximately equals the wavelength of the fastest-growing modes of MRI~\cite{Balbus1998} and $v_\mathrm{A} = B_0/\sqrt{4\pi\rho}$ is the Alfv\'en speed given by the initial magnetic field.

The computational domain was discretized into $N_x\times N_y\times N_z = N \times N\times N/2$ grid points.
We set $N$ at 256 initially and gradually increased it to 8192 after nonlinear saturation.
Each time $N$ was increased, we continued the simulation until the spectral shapes near the dissipation scale did not change before $N$ was increased again.
When $N$ was increased to 8192, the simulation was continued for a duration exceeding 100 $\Omega^{-1}$ from the initial time.
Thus, the MRI turbulence near the injection scale was sufficiently developed before we start the highest resolution run. 

As the numerical resolution increases, the simulation timestep must be decreased in order to satisfy the Courant–Friedrichs–Lewy condition.
We only computed for $\simeq 0.8\Omega^{-1}$ after $N$ increased from 4096 to 8192, and it was impossible to execute the highest resolution run over multiple eddy-turn-over time even though we exhausted $\simeq112$ million CPU hours.
However, we found that the spectral shapes at small scales ($k > 10$)  do not depend on time, as we can see in Fig.~S2, which shows the time history of spectra during the last $0.6\Omega^{-1}$ after the resolution increased (i.e., at $t = 0$ in Fig.~S3, approximately $0.2\Omega^{-1}$ passed after $N$ was increased).

To broaden the inertial range, we used fourth-order hyperviscosity $\nu_\mathrm{h}\nabla^4\mathbf{u}$ and hyperresistivity $\eta_\mathrm{h}\nabla^4\mathbf{B}$ to terminate the cascade.
Magnetic Prandtl number $\mathrm{Pm} = \nu_\mathrm{h}/\eta_\mathrm{h}$ was set to unity.
We found that when the order of the hyper-dissipation was eight, there appeared an unphysical roll up in the kinetic-energy spectrum.

\paragraph*{Shell-to-shell energy transfer function and cross-scale energy flux}
We first introduce the filtering in the wavenumber shell $|\mathbf{k}| = K$,
%----------------------------------------------------------------% 
\begin{align}
  \mathbf{u}_{K}(\mathbf{x}) = \sum_{K \le |\mathbf{k}| < K+1} \hat{\mathbf{u}}_\mathbf{k} \mathrm{e}^{\mathrm{i} \mathbf{k} \cdot \mathbf{x}}, \quad \mathbf{B}_{K}(\mathbf{x}) = \sum_{K \le |\mathbf{k}| < K+1} \hat{\mathbf{B}}_\mathbf{k} \mathrm{e}^{\mathrm{i} \mathbf{k} \cdot \mathbf{x}},
\end{align}
%----------------------------------------------------------------% 
where $\hat{\mathbf{u}}_\mathbf{k}$ and $\hat{\mathbf{B}}_\mathbf{k}$ are the Fourier coefficients of $\mathbf{u}$ and $\mathbf{B}$, respectively.
Then, the energy transfer between the field $f$ in the shell $Q \le |\mathbf{k}| < Q + 1$ to the field $g$ in the shell $K \le |\mathbf{k}| < K + 1$, denoted by $\mathcal{T}_{fg}(Q, K)$, is calculated as~\cite{Verma2004,Debliquy2005,Alexakis2005,Grete2017}
%----------------------------------------------------------------% 
\begin{align}
  \mathcal{T}_{uu}(Q, K) =& -\int\mathrm{d}^3\mathbf{r} \left[\mathbf{u}_K \cdot \left(\mathbf{u} \cdot \nabla \mathbf{u}_{Q} \right)\right] \\
  \mathcal{T}_{BB}(Q, K) =& -\int\mathrm{d}^3\mathbf{r} \left[\mathbf{B}_K \cdot \left(\mathbf{u} \cdot \nabla \mathbf{B}_{Q} \right)\right] \\
  \mathcal{T}_{uB}(Q, K) =&  \int\mathrm{d}^3\mathbf{r} \left[\mathbf{B}_K \cdot \left(\mathbf{B} \cdot \nabla \mathbf{u}_{Q} \right)\right] \\
  \mathcal{T}_{Bu}(Q, K) =&  \int\mathrm{d}^3\mathbf{r} \left[\mathbf{u}_K \cdot \left(\mathbf{B} \cdot \nabla \mathbf{B}_{Q} \right)\right].
\end{align}
%----------------------------------------------------------------% 
Integrating by parts, we obtain the identity $\mathcal{T}_{fg}(Q, K) = -\mathcal{T}_{gf}(K, Q)$.
The cross-scale energy flux across the wavenumber shell $|\mathbf{k}| = K$ is defined as the transfer from all scales larger than $K$ to all scales smaller than $K$.
It is given by~\cite{Grete2017}
%----------------------------------------------------------------% 
\begin{align}
  \Pi^{f^<}_{g^{>}}(k) = \sum_{Q < k}\sum_{K > k} \mathcal{T}_{fg}(Q, K).
\end{align}
\bibliography{references}

\begin{thebibliography}{10}

\bibitem{Balbus1991}
S.~A. {Balbus}, J.~F. {Hawley}, {\it {A Powerful Local Shear Instability in
  Weakly Magnetized Disks. I. Linear Analysis}\/}, {\it Astrophys.~J.\/} {\bf
  376}, 214-222 (1991).

\bibitem{Quataert1998}
E.~{Quataert}, {\it {Particle heating by Alfv{\'e}nic turbulence in hot
  accretion flows}\/}, {\it Astrophys.~J.\/} {\bf 500}, 978-991 (1998).

\bibitem{Quataert1999}
E.~{Quataert}, A.~{Gruzinov}, {\it {Turbulence and particle heating in
  advection-dominated accretion flows}\/}, {\it Astrophys.~J.\/} {\bf 520},
  248-255 (1999).

\bibitem{Kimura2016}
S.~S. {Kimura}, K.~{Toma}, T.~K. {Suzuki}, S.-i. {Inutsuka}, {\it {Stochastic
  Particle Acceleration in Turbulence Generated by Magnetorotational
  Instability}\/}, {\it Astrophys.~J.\/} {\bf 822}, 88 (2016).

\bibitem{Kimura2019}
S.~S. {Kimura}, K.~{Tomida}, K.~{Murase}, {\it {Acceleration and escape
  processes of high-energy particles in turbulence inside hot accretion
  flows}\/}, {\it Mon.\ Not.\ R.\ Astron.\ Soc.\/} {\bf 485}, 163-178 (2019).

\bibitem{Sun2021}
X.~{Sun}, X.-N. {Bai}, {\it {Particle diffusion and acceleration in
  magnetorotational instability turbulence}\/}, {\it Mon.\ Not.\ R.\ Astron.\
  Soc.\/} {\bf 506}, 1128-1147 (2021).

\bibitem{Stawarz2008}
{\L}.~{Stawarz}, V.~{Petrosian}, {\it {On the Momentum Diffusion of Radiating
  Ultrarelativistic Electrons in a Turbulent Magnetic Field}\/}, {\it
  Astrophys.~J.\/} {\bf 681}, 1725-1744 (2008).

\bibitem{Schekochihin2009}
A.~A. {Schekochihin}, {\it et~al.\/}, {\it {Astrophysical gyrokinetics: kinetic
  and fluid turbulent cascades in magnetized weakly collisional plasmas}\/},
  {\it \apjs\/} {\bf 182}, 310-377 (2009).

\bibitem{Kawazura2019}
Y.~{Kawazura}, M.~{Barnes}, A.~A. {Schekochihin}, {\it {Thermal
  disequilibration of ions and electrons by collisionless plasma
  turbulence}\/}, {\it Proc.\ Nat.\ Acad.\ Sci.\/} {\bf 116}, 771-776 (2019).

\bibitem{Kawazura2020}
Y.~{Kawazura}, {\it et~al.\/}, {\it {Ion versus electron heating in
  compressively driven astrophysical gyrokinetic turbulence}\/}, {\it Phys.\
  Rev.\ X\/} {\bf 10}, 041050 (2020).

\bibitem{Hawley1995}
J.~F. {Hawley}, C.~F. {Gammie}, S.~A. {Balbus}, {\it {Local three-dimensional
  magnetohydrodynamic simulations of accretion disks}\/}, {\it Astrophys.~J.\/}
  {\bf 440}, 742-763 (1995).

\bibitem{Fromang2007a}
S.~{Fromang}, J.~{Papaloizou}, {\it {MHD simulations of the magnetorotational
  instability in a shearing box with zero net flux. I. The issue of
  convergence}\/}, {\it Astron.\ Astrophys.\/} {\bf 476}, 1113-1122 (2007).

\bibitem{Fromang2007b}
S.~{Fromang}, J.~{Papaloizou}, G.~{Lesur}, T.~{Heinemann}, {\it {MHD
  simulations of the magnetorotational instability in a shearing box with zero
  net flux. II. The effect of transport coefficients}\/}, {\it Astron.\
  Astrophys.\/} {\bf 476}, 1123-1132 (2007).

\bibitem{Lesur2011}
G.~{Lesur}, P.~Y. {Longaretti}, {\it {Non-linear energy transfers in accretion
  discs MRI turbulence. I. Net vertical field case}\/}, {\it Astron.\
  Astrophys.\/} {\bf 528}, A17 (2011).

\bibitem{Walker2016}
J.~{Walker}, G.~{Lesur}, S.~{Boldyrev}, {\it {On the nature of magnetic
  turbulence in rotating, shearing flows}\/}, {\it Mon.\ Not.\ R.\ Astron.\
  Soc.\/} {\bf 457}, L39-L43 (2016).

\bibitem{Hirai2018}
K.~{Hirai}, Y.~{Katoh}, N.~{Terada}, S.~{Kawai}, {\it {Study of the transition
  from MRI to magnetic turbulence via parasitic instability by a high-order MHD
  simulation code}\/}, {\it Astrophys.~J.\/} {\bf 853}, 174 (2018).

\bibitem{Goldreich1995}
P.~{Goldreich}, S.~{Sridhar}, {\it {Toward a theory of interstellar turbulence.
  2: Strong Alfv\'enic turbulence}\/}, {\it Astrophys.~J.\/} {\bf 438}, 763-775
  (1995).

\bibitem{Boldyrev2006}
S.~{Boldyrev}, {\it {Spectrum of magnetohydrodynamic turbulence}\/}, {\it
  Phys.\ Rev.\ Lett.\/} {\bf 96}, 115002 (2006).

\bibitem{Zhdankin2017b}
V.~{Zhdankin}, J.~{Walker}, S.~{Boldyrev}, G.~{Lesur}, {\it {Universal
  small-scale structure in turbulence driven by magnetorotational
  instability}\/}, {\it Mon.\ Not.\ R.\ Astron.\ Soc.\/} {\bf 467}, 3620-3627
  (2017).

\bibitem{Cho2002}
J.~{Cho}, A.~{Lazarian}, {\it {Compressible sub-Alfv{\'e}nic MHD turbulence in
  low-{\ensuremath{\beta}} plasmas}\/}, {\it Phys.\ Rev.\ Lett.\/} {\bf 88},
  245001 (2002).

\bibitem{Cho2003}
J.~{Cho}, A.~{Lazarian}, {\it {Compressible magnetohydrodynamic turbulence:
  mode coupling, scaling relations, anisotropy, viscosity-damped regime and
  astrophysical implications}\/}, {\it Mon.\ Not.\ R.\ Astron.\ Soc.\/} {\bf
  345}, 325-339 (2003).

\bibitem{Makwana2020}
K.~D. {Makwana}, H.~{Yan}, {\it {Properties of magnetohydrodynamic modes in
  compressively driven plasma turbulence}\/}, {\it Phys.\ Rev.\ X\/} {\bf 10},
  031021 (2020).

\bibitem{Teraki2019}
Y.~{Teraki}, K.~{Asano}, {\it {Particle Energy Diffusion in Linear
  Magnetohydrodynamic Waves}\/}, {\it \apj\/} {\bf 877}, 71 (2019).

\bibitem{Lynn2014}
J.~W. {Lynn}, E.~{Quataert}, B.~D.~G. {Chandran}, I.~J. {Parrish}, {\it
  {Acceleration of Relativistic Electrons by Magnetohydrodynamic Turbulence:
  Implications for Non-thermal Emission from Black Hole Accretion Disks}\/},
  {\it \apj\/} {\bf 791}, 71 (2014).

\bibitem{Murase2020}
K.~{Murase}, S.~S. {Kimura}, P.~{M{\'e}sz{\'a}ros}, {\it {Hidden Cores of
  Active Galactic Nuclei as the Origin of Medium-Energy Neutrinos: Critical
  Tests with the MeV Gamma-Ray Connection}\/}, {\it \prl\/} {\bf 125}, 011101
  (2020).

\bibitem{Kimura2021}
S.~S. {Kimura}, K.~{Murase}, P.~{M{\'e}sz{\'a}ros}, {\it {Soft gamma rays from
  low accreting supermassive black holes and connection to energetic
  neutrinos}\/}, {\it Nat. Commun.\/} {\bf 12}, 5615 (2021).

\bibitem{IceCube2022}
{IceCube Collaboration}, {\it et~al.\/}, {\it {Evidence for neutrino emission
  from the nearby active galaxy NGC 1068}\/}, {\it Science\/} {\bf 378},
  538-543 (2022).

\bibitem{Kawazura2022a}
Y.~{Kawazura}, A.~A. {Schekochihin}, M.~{Barnes}, W.~{Dorland}, S.~A. {Balbus},
  {\it {Energy partition between Alfv{\'e}nic and compressive fluctuations in
  magnetorotational turbulence with near-azimuthal mean magnetic field}\/},
  {\it J.\ Plasma Phys.\/} {\bf 88}, 905880311 (2022).

\bibitem{Kunz2016}
M.~W. {Kunz}, J.~M. {Stone}, E.~{Quataert}, {\it {Magnetorotational Turbulence
  and Dynamo in a Collisionless Plasma}\/}, {\it Phys.\ Rev.\ Lett.\/} {\bf
  117}, 235101 (2016).

\bibitem{Boldyrev2011}
S.~{Boldyrev}, J.~C. {Perez}, J.~E. {Borovsky}, J.~J. {Podesta}, {\it {Spectral
  Scaling Laws in Magnetohydrodynamic Turbulence Simulations and in the Solar
  Wind}\/}, {\it \apjl\/} {\bf 741}, L19 (2011).

\bibitem{Verma2004}
M.~K. {Verma}, {\it {Statistical theory of magnetohydrodynamic turbulence:
  recent results}\/}, {\it Phys.\ Rep.\/} {\bf 401}, 229-380 (2004).

\bibitem{Debliquy2005}
O.~{Debliquy}, M.~K. {Verma}, D.~{Carati}, {\it {Energy fluxes and
  shell-to-shell transfers in three-dimensional decaying magnetohydrodynamic
  turbulence}\/}, {\it Phys.\ Plasmas\/} {\bf 12}, 042309 (2005).

\bibitem{Alexakis2005}
A.~{Alexakis}, P.~D. {Mininni}, A.~{Pouquet}, {\it {Shell-to-shell energy
  transfer in magnetohydrodynamics. I. Steady state turbulence}\/}, {\it Phys.\
  Rev.\ E\/} {\bf 72}, 046301 (2005).

\bibitem{Grete2017}
P.~{Grete}, B.~W. {O'Shea}, K.~{Beckwith}, W.~{Schmidt}, A.~{Christlieb}, {\it
  {Energy transfer in compressible magnetohydrodynamic turbulence}\/}, {\it
  Phys.\ Plasmas\/} {\bf 24}, 092311 (2017).

\bibitem{Cho2004}
J.~{Cho}, A.~{Lazarian}, {\it {The Anisotropy of Electron Magnetohydrodynamic
  Turbulence}\/}, {\it Astrophys.~J.\/} {\bf 615}, L41-L44 (2004).

\bibitem{Satapathy2024}
K.~{Satapathy}, D.~{Psaltis}, F.~{{\"O}zel}, {\it {The origin of the
  Slow-to-Alfv{\'e}n Wave Cascade Power Ratio and its Implications for Particle
  Heating in Accretion Flows}\/}, {\it arXiv e-prints\/} p. arXiv:2402.14089
  (2024).

\bibitem{Loren2024}
L.~E. {Held}, G.~{Mamatsashvili}, M.~E. {Pessah}, {\it {MRI turbulence in
  vertically stratified accretion discs at large magnetic Prandtl numbers}\/},
  {\it Mon.\ Not.\ R.\ Astron.\ Soc.\/} {\bf 530}, 2232-2250 (2024).

\bibitem{EHT2019e}
{EHT Collaboration}, {\it {First M87 Event Horizon Telescope Results. V.
  Physical origin of the asymmetric ring}\/}, {\it Astrophys.~J.\/} {\bf 875},
  L5 (2019).

\bibitem{Sharma2007}
P.~{Sharma}, E.~{Quataert}, G.~W. {Hammett}, J.~M. {Stone}, {\it {Electron
  heating in hot accretion flows}\/}, {\it Astrophys.~J.\/} {\bf 667}, 714-723
  (2007).

\bibitem{Kempski2019}
P.~{Kempski}, E.~{Quataert}, J.~{Squire}, M.~W. {Kunz}, {\it {Shearing-box
  simulations of MRI-driven turbulence in weakly collisional accretion
  discs}\/}, {\it Mon.\ Not.\ R.\ Astron.\ Soc.\/} {\bf 486}, 4013-4029 (2019).

\bibitem{Satapathy2023}
K.~{Satapathy}, D.~{Psaltis}, F.~{{\"O}zel}, {\it {Global Electron
  Thermodynamics in Radiatively Inefficient Accretion Flows}\/}, {\it \apj\/}
  {\bf 955}, 47 (2023).

\bibitem{EHT2021}
{Event Horizon Telescope Collaboration}, {\it et~al.\/}, {\it {First M87 Event
  Horizon Telescope Results. VIII. Magnetic Field Structure near The Event
  Horizon}\/}, {\it \apjl\/} {\bf 910}, L13 (2021).

\bibitem{EHT2022}
{Event Horizon Telescope Collaboration}, {\it et~al.\/}, {\it {First
  Sagittarius A* Event Horizon Telescope Results. V. Testing Astrophysical
  Models of the Galactic Center Black Hole}\/}, {\it \apjl\/} {\bf 930}, L16
  (2022).

\bibitem{Narayan2003}
R.~{Narayan}, I.~V. {Igumenshchev}, M.~A. {Abramowicz}, {\it {Magnetically
  Arrested Disk: an Energetically Efficient Accretion Flow}\/}, {\it Publ.
  Astron. Soc. Japan\/} {\bf 55}, L69-L72 (2003).

\bibitem{Jacquemin-Ide2023}
J.~{Jacquemin-Ide}, F.~{Rincon}, A.~{Tchekhovskoy}, M.~{Liska}, {\it
  {Magnetorotational dynamo can generate large-scale vertical magnetic fields
  in 3D GRMHD simulations of accreting black holes}\/}, {\it arXiv e-prints\/}
  p. arXiv:2311.00034 (2023).

\bibitem{Begelman2022}
M.~C. {Begelman}, N.~{Scepi}, J.~{Dexter}, {\it {What really makes an accretion
  disc MAD}\/}, {\it \mnras\/} {\bf 511}, 2040-2051 (2022).

\bibitem{Hoshino2013}
M.~{Hoshino}, {\it {Particle Acceleration during Magnetorotational Instability
  in a Collisionless Accretion Disk}\/}, {\it Astrophys.~J.\/} {\bf 773}, 118
  (2013).

\bibitem{Hoshino2015}
M.~{Hoshino}, {\it {Angular Momentum Transport and Particle Acceleration During
  Magnetorotational Instability in a Kinetic Accretion Disk}\/}, {\it Phys.\
  Rev.\ Lett.\/} {\bf 114}, 061101 (2015).

\bibitem{Bacchini2022}
F.~{Bacchini}, {\it et~al.\/}, {\it {Fully Kinetic Shearing-box Simulations of
  Magnetorotational Turbulence in 2D and 3D. I. Pair Plasmas}\/}, {\it
  Astrophys.~J.\/} {\bf 938}, 86 (2022).

\bibitem{Comisso2018}
L.~{Comisso}, L.~{Sironi}, {\it {Particle Acceleration in Relativistic Plasma
  Turbulence}\/}, {\it \prl\/} {\bf 121}, 255101 (2018).

\bibitem{Sharma2006}
P.~{Sharma}, G.~W. {Hammett}, E.~{Quataert}, J.~M. {Stone}, {\it {Shearing Box
  Simulations of the MRI in a Collisionless Plasma}\/}, {\it Astrophys.~J.\/}
  {\bf 637}, 952-967 (2006).

\bibitem{Riquelme2012}
M.~A. {Riquelme}, E.~{Quataert}, P.~{Sharma}, A.~{Spitkovsky}, {\it {Local
  Two-dimensional Particle-in-cell Simulations of the Collisionless
  Magnetorotational Instability}\/}, {\it Astrophys.~J.\/} {\bf 755}, 50
  (2012).

\bibitem{Foucart2016}
F.~{Foucart}, M.~{Chandra}, C.~F. {Gammie}, E.~{Quataert}, {\it {Evolution of
  accretion discs around a kerr black hole using extended
  magnetohydrodynamics}\/}, {\it Mon.\ Not.\ R.\ Astron.\ Soc.\/} {\bf 456},
  1332-1345 (2016).

\bibitem{Foucart2017}
F.~{Foucart}, M.~{Chandra}, C.~F. {Gammie}, E.~{Quataert}, A.~{Tchekhovskoy},
  {\it {How important is non-ideal physics in simulations of sub-Eddington
  accretion on to spinning black holes?}\/}, {\it Mon.\ Not.\ R.\ Astron.\
  Soc.\/} {\bf 470}, 2240-2252 (2017).

\bibitem{Inchingolo2018}
G.~{Inchingolo}, T.~{Grismayer}, N.~F. {Loureiro}, R.~A. {Fonseca}, L.~O.
  {Silva}, {\it {Fully Kinetic Large-scale Simulations of the Collisionless
  Magnetorotational Instability}\/}, {\it Astrophys.~J.\/} {\bf 859}, 149
  (2018).

\bibitem{Bacchini2024}
F.~{Bacchini}, {\it et~al.\/}, {\it {Collisionless Magnetorotational Turbulence
  in Pair Plasmas: Steady-state Dynamics, Particle Acceleration, and Radiative
  Cooling}\/}, {\it arXiv e-prints\/} p. arXiv:2401.01399 (2024).

\bibitem{Kawazura2022b}
Y.~{Kawazura}, {\it {CALLIOPE: Pseudospectral Shearing Magnetohydrodynamics
  Code with a Pencil Decomposition}\/}, {\it Astrophys.~J.\/} {\bf 928}, 113
  (2022).

\bibitem{Kawazura2022d}
Y.~{Kawazura}, {\it {Integrating Factor Runge-Kutta Method in Shearing
  Coordinates}\/}, {\it J.\ Phys.\ Soc.\ Japan\/} {\bf 91}, 115002 (2022).

\bibitem{Rogallo1981}
R.~S. {Rogallo}, {Numerical experiments in homogeneous turbulence}, NASA
  STI/Recon Technical Report N (1981).

\bibitem{Umurhan2004}
O.~M. {Umurhan}, O.~{Regev}, {\it {Hydrodynamic stability of rotationally
  supported flows: Linear and nonlinear 2D shearing box results}\/}, {\it
  Astron.\ Astrophys.\/} {\bf 427}, 855-872 (2004).

\bibitem{Balbus1998}
S.~A. {Balbus}, J.~F. {Hawley}, {\it {Instability, turbulence, and enhanced
  transport in accretion disks}\/}, {\it Rev.\ Mod.\ Phys.\/} {\bf 70}, 1-53
  (1998).

\end{thebibliography}

\bibliographystyle{Science}

\subsection*{Acknowledgments}
YK thanks Hideaki Miura, Tomohiko Watanabe, and Jim Stone for fruitful comments on this study. 
Numerical computations reported here were carried out on Fugaku at the RIKEN Center for Computational Science (Project ID: hp220027 and hp230006), on ATERUI II at Center for Computational Astrophysics in National Astronomical Observatory of Japan, on Oakbridge-CX, Oakforest-PACS and Wisteria/BDEC-01 at the University of Tokyo, on Flow at Nagoya University, and on ITO at Kyushu University. 

\subsection*{Funding}
\begin{itemize}
  \setlength{\itemsep}{0cm}
  \setlength{\parskip}{0cm}
  \item[] YK is supported by JSPS KAKENHI grant No. 20K14509. SSK is supported by JSPS KAKENHI grant Nos. 22K14028, 21H04487, 23H04899 and the Tohoku Initiative for Fostering Global Researchers for Interdisciplinary Sciences (TI-FRIS) of MEXTs Strategic Professional Development Program for Young Researchers.
\end{itemize}

\subsection*{Author contributions}
\begin{itemize}
  \setlength{\itemsep}{0cm}
  \setlength{\parskip}{0cm}
  \item[] Conceptualization: YK and SSK
  \item[] Methodology: YK
  \item[] Investigation: YK and SSK
  \item[] Visualization: YK
  \item[] Funding acquisition: YK and SSK
  \item[] Writing – original draft: YK and SSK
  \item[] Writing – review \& editing: YK and SSK
\end{itemize}

\subsection*{Competing interests}
\begin{itemize}
  \setlength{\itemsep}{0cm}
  \setlength{\parskip}{0cm}
  \item[] The authors declare that they have no competing interests.
\end{itemize}

\subsection*{Data and materials availability}
\begin{itemize}
  \setlength{\itemsep}{0cm}
  \setlength{\parskip}{0cm}
  \item[] All data needed to evaluate the conclusions in the paper are present in the paper and/or the Supplementary Materials.
\end{itemize}

%Here you should list the contents of your Supplementary Materials -- below is an example. 
%You should include a list of Supplementary figures, Tables, and any references that appear only in the SM. 
%Note that the reference numbering continues from the main text to the SM.
% In the example below, Refs. 4-10 were cited only in the SM.     
\section*{Supplementary materials}
Figs. S1 to S3.

% For your review copy (i.e., the file you initially send in for
% evaluation), you can use the {figure} environment and the
% \includegraphics command to stream your figures into the text, placing
% all figures at the end.  For the final, revised manuscript for
% acceptance and production, however, PostScript or other graphics
% should not be streamed into your compliled file.  Instead, set
% captions as simple paragraphs (with a \noindent tag), setting them
% off from the rest of the text with a \clearpage as shown  below, and
% submit figures as separate files according to the Art Department's
% instructions.

\clearpage

%----------------------------------------------------------------% 

%----------------------------------------------------------------% 

\clearpage

\begin{center}
{\Large Supplementary Materials for} \\
\textbf{\Large Inertial range of magnetorotational turbulence} \\
\vskip1em
{\large Yohei Kawazura$^*$, Shigeo S. Kimura}\\
\vskip1em
{\large *Correspondence to: kawazura@a.utsunomiya-u.ac.jp}\\
\end{center}

\textbf{This PDF file includes:}
\vspace{-1em}
\begin{itemize}
  \setlength{\itemsep}{0cm}
  \setlength{\parskip}{0cm}
  \item[] Figs. S1 to S3
\end{itemize}

\clearpage

\renewcommand\thefigure{S\arabic{figure}}
\setcounter{figure}{0}
%----------------------------------------------------------------% 
\begin{figure}
  \begin{center}
    \includegraphics*[width=1.0\textwidth]{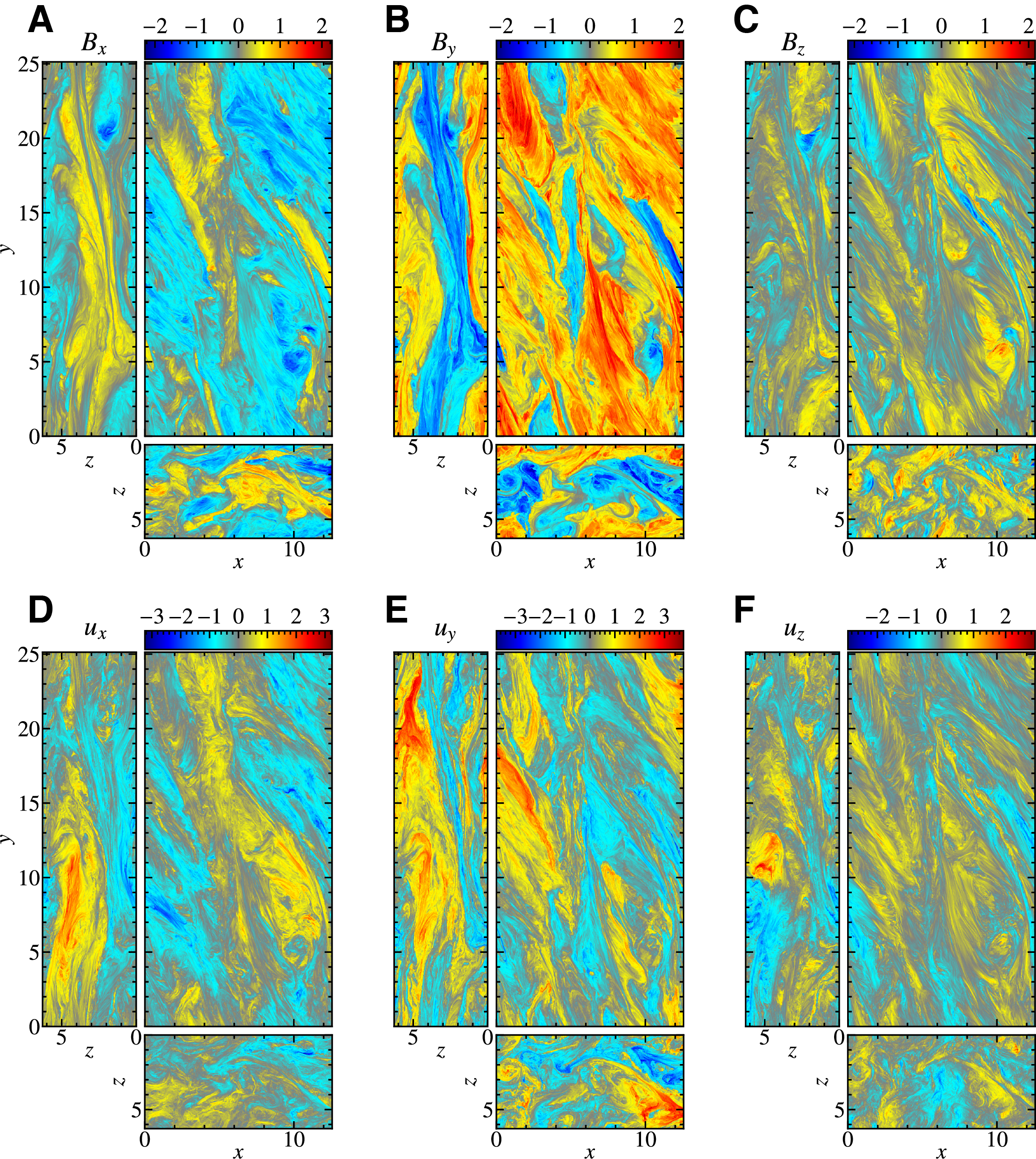}
  \end{center}
  \caption{
    \textbf{The spatial distribution of the vector components of magnetic and flow fields.}
    \textbf{(A)}, \textbf{(B)}, and \textbf{(C)} for the magnetic fields $B_x$, $B_y$, and $B_z$ and \textbf{(D)}, \textbf{(E)}, and \textbf{(F)} for the flow fields $u_x$, $u_y$, and $u_z$.
    The snapshot was taken at the same time as in Fig.~\ref{f:snapshot}.
    One finds that only $B_y$ has intense large-scale structures, which are supposed to be created by the $\Omega$ effect and inverse cascade, as mentioned in the main text.
  }
\end{figure}
%----------------------------------------------------------------% 
\begin{figure}
  \begin{center}
    \includegraphics*[width=1.0\textwidth]{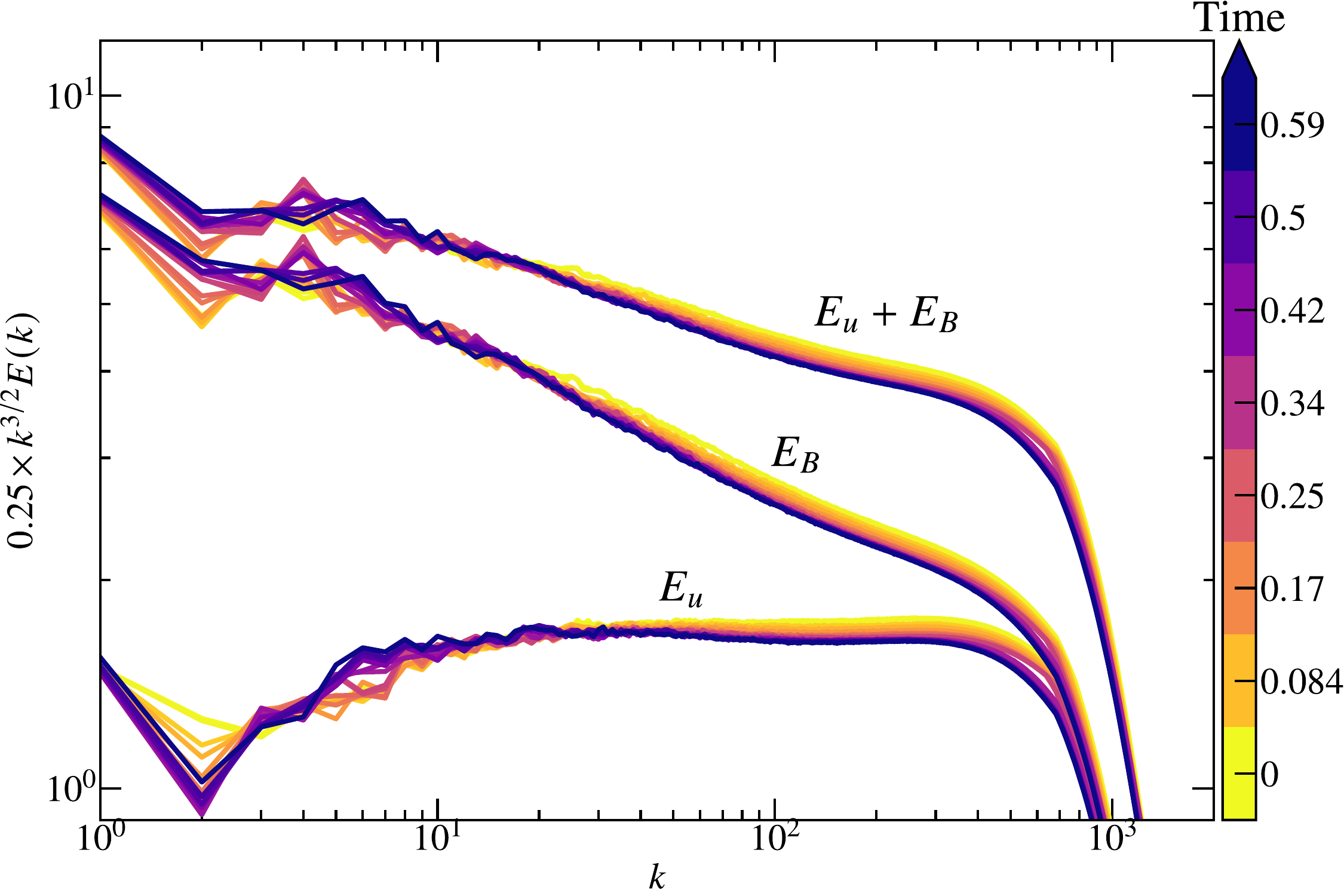}
  \end{center}
  \caption{
    \textbf{Time evolution of spectra.} The evolution is displayed from $t = 0$ to $0.6\Omega^{-1}$ where $t = 0$ corresponds to the time when $0.2\Omega^{-1}$ passed after the resolution was increased to $N = 8192$.
  }
\end{figure}
%----------------------------------------------------------------% 
\begin{figure}
  \begin{center}
    \includegraphics*[width=1.0\textwidth]{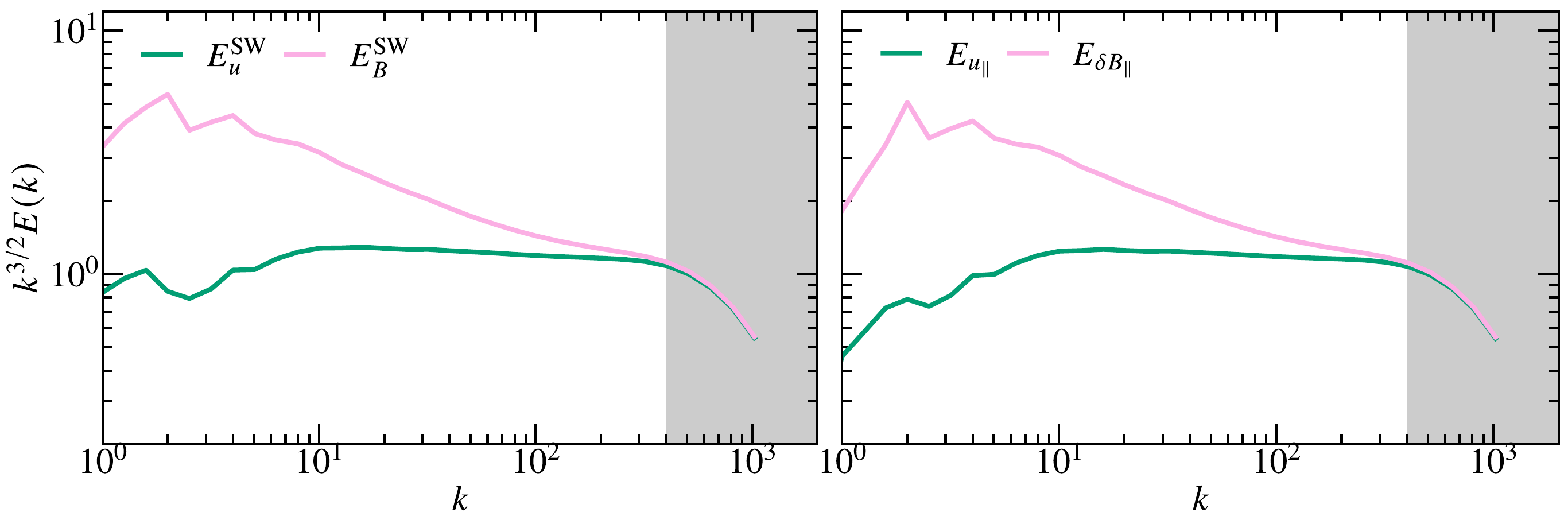}
  \end{center}
  \caption{
    \textbf{Comparison between the spectra of slow-magnetosonic fluctuations and parallel fluctuations.}
    (Left) The spectrum of magnetic and kinetic energy of slow-magnetosonic waves $E_u^\mathrm{SW} = (1 + k_\|^2/k_\perp^2)u_\|^2/2$ and $E_B^\mathrm{SW} = (1 + k_\|^2/k_\perp^2)\delta B_\|^2/2$, respectively, and (Right) the spectrum of $u_\|^2/2$ and $\delta B_\|^2/2$ (the same as Fig.~\ref{f:SW-AW}A). The only difference is at low-$k$ region were the anisotropy $k_\|/k_\perp \ll 1$ is not yet developed.
  }
\end{figure}
%----------------------------------------------------------------% 

\end{document}